\documentstyle[twocolumn,epsf]{jpsj}

\def\runtitle{
Spin Splitting in de Haas-van Alp-hen Oscillation
in Two-Dimensional Two-Band Systems 
}
\def\runauthor{Keita {\sc Kishigi}, 
Yasumasa {\sc Hasegawa} and 
Mitake {\sc Miyazaki}
}

\title{
Spin Splitting in de Haas-van Alphen Oscillation
in Two-Dimensional Two-Band Systems 
}

\author
{
Keita {\sc Kishigi}, 
Yasumasa {\sc Hasegawa} and 
Mitake {\sc Miyazaki}
}

\inst
{
Faculty of Science, Himeji Institute of Technology, Akou-gun, Hyogo
678-1297, Japan 
}

\recdate
{
\today
}

\abst
{
We study the effects of the Zeeman term on 
the de Haas van Alphen (dHvA) oscillation 
in two-dimensional two-band systems. 
We found that the Fourier transform 
amplitudes of the oscillations are not described by the spin 
reduction factor in the Lifshitz-Kosevich formalism 
in two-dimensional systems. 
The anomalous dependence on the effective $g$-factor 
can be observed by tilting-angle dependence of the 
dHvA oscillation 
in quasi-two-dimensional organic conductors and Sr$_2$RuO$_4$. 
}

\kword
{dHvA Oscillation, Sr$_2$RuO$_4$, Quasi-Two-Dimensional
Organic Conductors, Zeeman effect, Magnetic Breakdown 
}

\begin{document}
\sloppy
\maketitle
The magnetization and 
magnetoresistance oscillate 
as a function of the inverse of the 
magnetic field ($H$) 
with the period 
proportional to the extreme area ($f$) of 
the closed orbit 
on the Fermi surface. 
These phenomena are known as 
the Shubnikov-de Haas (SdH) oscillation and 
the de Haas-van Alphen (dHvA) oscillation. 
The dHvA oscillation can be 
described in the semiclassical theory 
as Lifshitz and Kosevich (LK) formula\cite{LK,Shoenberg84}. 

In this study, we consider the 
system at the clean limit and zero temperature. 
In the two-dimensional single-band system, 
the magnetization, $M^{s}(\mu, H)$, 
for the fixed chemical potential ($\mu$) 
is given by\cite{LK,Shoenberg84} 
\begin{eqnarray}
M^{s}(\mu,H)&=&-\sum_{p=1}^{\infty}R^{\mu}_{p}\frac{1}{p}\sin p(\frac{f}{H}-\pi),\\ 
R_{p}^{\mu}&=&\cos (\frac{\pi}{2}pg\frac{m}{m_{0}}), 
\end{eqnarray}
where $R_{p}^{\mu}$ is 
the reduction factor 
due to electron spin, 
$g\simeq2$ is the $g$-factor, $m$ is 
the cyclotron effective mass and $m_0$ is the free electron mass. 
The spin factor is caused by the phase shift in the oscillation 
for up and down spins. 
When the Zeeman splitting becomes 
a half of the Landau level spacing, 
the amplitude of the oscillation of 
the fundamental frequency ($p=1$) becomes zero. 
Eq. (1) is called the LK formula 
\cite{Shoenberg84}. 
The experiments are performed in the 
multi energy-band system with the fixed electron 
number ($N$), where the chemical potential 
varies as a function of the magnetic field. 
However, the oscillation of the chemical potential is 
very small when the Fermi surface has a 
three-dimensional shape. 
Then we can apply eq. (1) or the superposition of eq. (1) with 
some frequencies even for the system with fixed 
electron number.

On the other hand, if the system is two-dimensional or 
the interlayer coupling is much smaller 
than the Landau level spacings (this 
condition is satisfied in 
the quasi-two-dimensional system in the 
strong magnetic field), 
the oscillation of the chemical potential plays important roll. 
The magnetization 
for the fixed electron number 
is given by\cite{LK,Shoenberg84} 
\begin{eqnarray}
M^{s}(N,H)&=&\sum_{p=1}^{\infty}R^{N}_{p}\frac{1}{p}\sin p\frac{f}{H}, 
\end{eqnarray}
in the two-dimensional single-band system. 
The spin reduction factor 
in this case is\cite{kishigispin,nakanospin} 
\begin{eqnarray}
|R_{p}^{N}|^2=\left\{\begin{array}{cc}|1
-2(gm/2m_0-[gm/2m_0])|^2 & 
{\rm for\
odd\ }p\\
1&{\rm for\ even\ }p, \end{array}\right. 
\end{eqnarray}
where $[gm/2m_0]$ is an 
integer part of $gm/2m_0$. 
The spin reduction factor, $R_{p}^{N}$, is the periodic 
function of $gm/2m_0$ but is not the 
cosine function. 

If the magnetic field is tilted by $\theta$ from 
the z-axis, 
the spin reduction factor is given by 
replacing $g$ with $g/\cos\theta$ in 
eqs. (2) and (4). 
The reduction factor 
for the $p$th harmonics, $R_{p}^{\mu}$, is zero when 
$\cos\theta=(pgm/m_0)/(2n+1)$ with integer $n$. 
When $\cos\theta=(gm/m_0)/(2n+1)$, $R_{p}^{N}$ for 
odd $p$ is zero. 
The tilting angle, $\theta$, when $R_{p}^{\mu}$ or $R_{p}^{N}$ 
becomes zero 
is called spin splitting zero 
condition\cite{Shoenberg84}, which 
can be measured by 
tilting the magnetic field. 

When two energy bands exist in the two-dimensional system, 
the magnetization is not described by the 
superposition of the magnetization for 
the single band (eq. (3)) if the electron 
number is fixed
~\cite{kishigi,kishigi2,kishigi3,harrison,sandu,so,fortin,nakano,alex}. 
One of the authors\cite{kishigi,kishigi2,kishigi3}, 
Harrison et al.\cite{harrison}, Sandu et al.\cite{sandu}, 
Han et al.\cite{so} and Fortin et al.\cite{fortin} treat 
the magnetic breakdown model 
as shown in Fig. 1(a), 
where the $\beta$ orbit is obtained by the tunneling through the 
zone gap. 
Nakano\cite{nakano} and 
Alexandrov and Bratkovsky\cite{alex} have calculated 
the dHvA oscillation 
by using the independent two-band model, 
in which there exist two closed orbits 
($\alpha$ and $\beta$ orbits) as shown in Fig. 1 (b). 
From these calculations,~\cite{kishigi,kishigi2,kishigi3,harrison,sandu,so,fortin,nakano,alex} 
 it has been shown that 
there exist 
frequencies 
in the magnetization ($M(N,H)$), which 
are forbidden in the semiclassical theory. 
One of these forbidden frequencies corresponds to the area 
of the $\beta$-$\alpha$ orbit. 
On the other hand, 
the forbidden oscillations do not appear in 
$M(\mu,H)$ 
when $\mu$ is fixed
\cite{kishigi3,harrison,so,fortin,nakano,alex}. 
These forbidden oscillations are caused by the chemical potential 
oscillation.

The existence of the $\beta$-$\alpha$ oscillation 
in the magnetization has been observed experimentally 
in quasi-two-dimensional materials 
with two or three energy bands\cite{FS}, 
$\kappa$-(BEDT-TTF)$_2$Cu(NCS)$_2$\cite{Meyer,uji}, 
$\alpha$-(BEDT-TTF)$_2$KHg(SCN)$_4$\cite{honold} 
and Sr$_2$RuO$_4$\cite{settai}. 

\begin{figure}
\leavevmode
\epsfxsize=6cm
\epsfbox{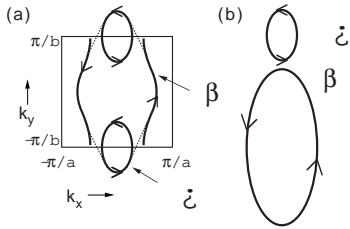}
\caption{(a) Fermi surface in the magnetic breakdown system. 
(b) Fermi surface in the two-band system. 
}
\label{fig:1}
\end{figure}

The spin was neglected to simplify the calculation
~\cite{kishigi,kishigi2,kishigi3,harrison,sandu,so,fortin,nakano,alex}.  
In order to study how the spin affects the 
``forbidden'' 
$\beta$-$\alpha$ oscillation 
and quantum magnetic oscillations ($\alpha$, $\beta$, 
$\beta$+$\alpha$, $2\alpha$ oscillations, etc.) 
in the two-dimensional magnetic breakdown systems, 
we have studied $M(N,H)$ in the tight-binding model 
quantum-mechanically\cite{kishigispin}. 
This model can be applied to the quasi-two-dimensional 
organic conductors. 
We found that 
the Fourier transform amplitudes (FTAs) of 
the $\beta$$+$$\alpha$ oscillation is {\it enhanced} 
due to the Zeeman effect, 
which is quite different from the 
spin reduction factors, 
eq. (2) or eq. (4). 
However, as far as we know, the effect of the Zeeman term in the 
two-dimensional two-band model has never been 
discussed systematically.

In this letter, 
we calculate $M(N,H)$ at $T=0$ for 
the two-dimensional two-band model, 
where the magnetic breakdown is neglected. 
In this case neither chemical potential nor electron number 
for each band and each spin are fixed. Our purpose is 
to make clear how the 
oscillation in $M(N,H)$ 
are modulated by an electron spin. 



\begin{figure}
\leavevmode
\epsfxsize=8.5cm
\epsfbox{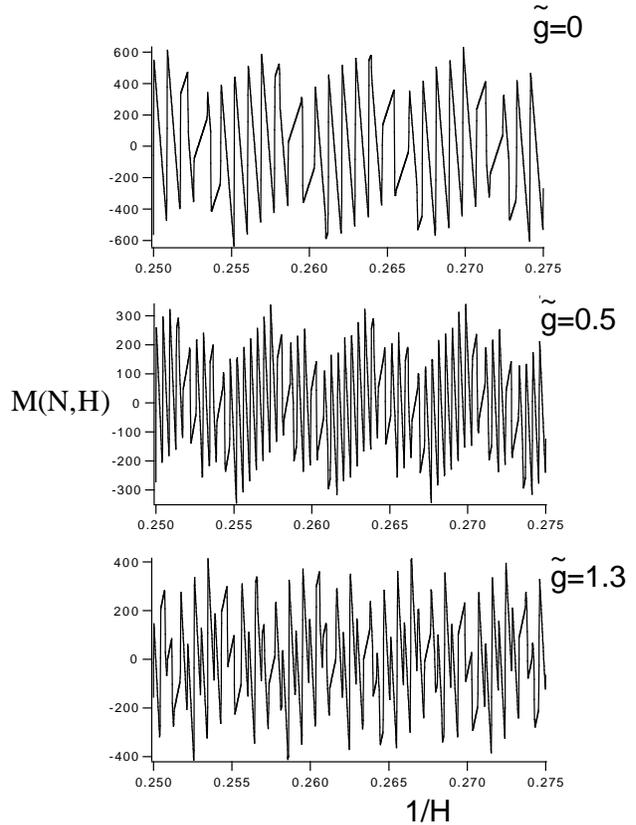}
\caption{
$M(N, H)$ 
as a function of $1/H$ when 
$\widetilde{g}=0, 1.0$ and $2.6$. 
}
\end{figure}

We study the model, whose energy 
at $H=0$ 
is given by 
\begin{eqnarray}
E_{i}({\bf k})=\frac{{\hbar}^{2} k^2_x}{2m^{x}_{i}}
+\frac{{\hbar}^2 k^2_y}{2m^{y}_{i}}+\epsilon^{b}_{i}, 
\end{eqnarray}
where $\epsilon^{b}_{i}$ is the bottom energy of the energy band 
and $i=\alpha$ and $\beta$ are the band index. 
The density of states for spin $\rho_{i}$ 
and the cyclotron effective mass $m_i$ are 
\begin{eqnarray}
\rho_{i}=Cm_i, m_i=\sqrt{m_i^x m_i^y}, 
\end{eqnarray}
where $C=S/2\pi\hbar^{2}$ and $S$ is the real 
space area of the two-dimensional system. 

When the magnetic field ($H_0$) is applied with the tilting angle 
$\theta$, the $z$-component of the magnetic field is 
$H=H_0\cos\theta$. 
The energy spectrum 
is quantized as 
\begin{eqnarray}
\epsilon_{i} (H,n,\sigma)=\hbar\omega_{i}
(n+\frac{1}{2}+\sigma(\frac{m_{i}}{m_{0}})\widetilde{g}_{i})
+\epsilon^{b}_{i}, 
\end{eqnarray}
where $n$ is Landau index, $\omega_{i}= eH/m_{i}c$, 
$\sigma=\pm\frac{1}{2}$ 
and $\widetilde{g}_{i}=g_{i}/2\cos\theta$. 
We take $g_{\alpha}=g_{\beta}$ 
and $\widetilde{g}_{i}=\widetilde{g}$, 
although $g_{\alpha}$ and $g_{\beta}$ are not necessarily the same 
due to the spin-orbit 
coupling. 
We set $e=\hbar =c=1$ and  
$C=1$.


When the electron number ($N$) is fixed, we have to 
calculate the Helmholtz free energy $F(N,H)$, which is given 
by 
\begin{eqnarray}
F(N,H)=\sum_{i}\rho_{i}\hbar\omega_{i}\sum_{n,\sigma}
\epsilon_{i}(H,n,\sigma), 
\end{eqnarray}
where the summations are performed for the filled 
and partially filled energy levels. 
The electron number, $N$, is decided 
by the chemical potential, $\mu (0)$, at $H=0$. 
The magnetization is obtained by 
\begin{eqnarray}
M(N, H)=-\partial F(N,H)/\partial H. 
\end{eqnarray}
We calculate $M(N, H)$ by changing 
the value of $\widetilde{g}$. 
When $\widetilde{g}=0$, $M(N, H)$ are 
reduced to the spinless model\cite{nakano}.

We show $M(N, H)$ 
for various values of $\widetilde{g}$ in Fig. 2, where 
$m_{\alpha}/m_{0}=0.5, m_{\beta}/m_{0}=1.0, \epsilon^{b}_{\alpha}=680, 
\epsilon^{b}_{\beta}=0$ and $\mu (0)= 1000$. 
We take $C=1$ and 
$0.25\leq 1/H\leq0.275$, which corresponds to about $8$ tesla 
and $\mu (0) \simeq0.2$ eV. 
In these parameters, the area of the $\alpha$ and $\beta$ 
orbits corresponds to $f_{\alpha}=160$ and 
$f_{\beta}=1000$.

We calculate the FTAs of 
$M(N,H)$, where 
there exit 
many combination frequencies, $\beta\pm\alpha$, 
$\beta\pm2\alpha$, $2\beta\pm\alpha$, $2\beta\pm2\alpha$, 
in addition to $\alpha$, $\beta$, $2\alpha$ and $2\beta$, 
which are shown in Fig. 3.


\begin{figure}
\leavevmode
\epsfxsize=7.5cm
\epsfbox{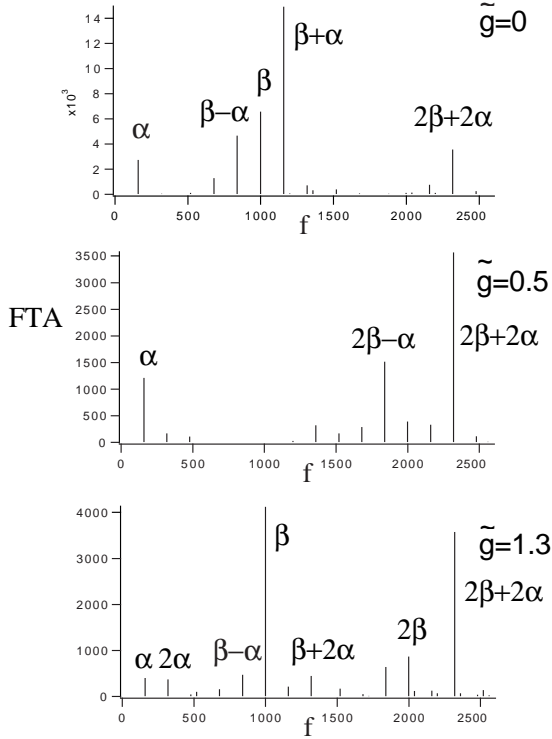}
\caption{The Fourier transform 
amplitudes of $M(N, H)$ in Fig. 2. 
}
\end{figure}



We show the FTAs of $M(N,H)$ as a 
function of $\widetilde{g}$ in Figs. 4 and 5. 
We also show FTAs for other 
parameters
($m_{\alpha}/m_{0}=0.65$, 
$m_{\beta}/m_{0}=1.0$, 
$\epsilon_{\alpha}^{b}=754$, 
$\epsilon_{\beta}^{b}=0$ and $\mu (0)=1000$) 
in Figs. 6 and 7. 



\begin{figure}
\leavevmode
\epsfxsize=7.5cm
\epsfbox{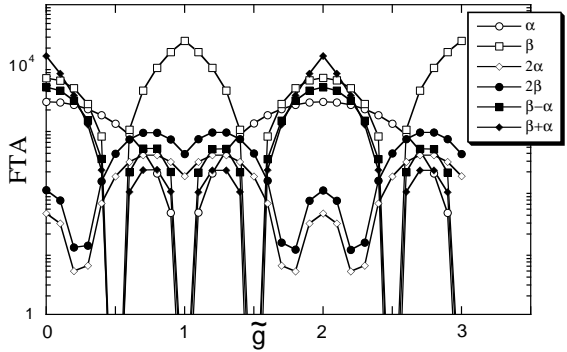}
\caption{The FTAs of 
some peaks ($f_\alpha$, $f_{2\alpha}$, $f_\beta$, $f_{2\beta}$, 
$f_{\beta-\alpha}$ and $f_{\beta+\alpha}$) in $M(N,H)$ 
as a function of $\widetilde{g}$ for 
$m_{\alpha}/m_{0}=0.5, m_{\beta}/m_{0}=1.0, \epsilon^{b}_{\alpha}=680$ and 
$\epsilon^{b}_{\beta}=0$. 
}
\end{figure}
\begin{figure}
\leavevmode
\epsfxsize=8cm
\epsfbox{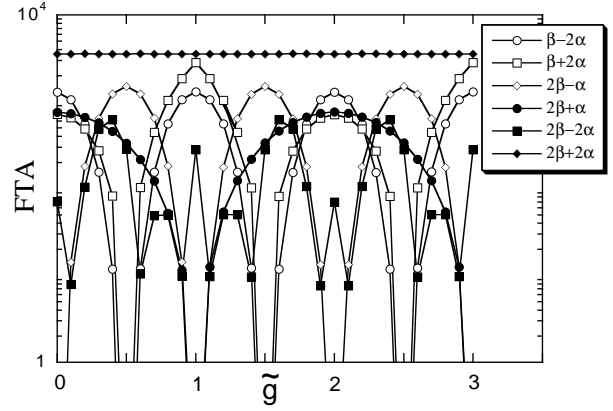}
\caption{The FTAs of 
some peaks ($f_{\beta-2\alpha}$, $f_{\beta+2\alpha}$, 
$f_{2\beta-\alpha}$, 
$f_{2\beta+\alpha}$, 
$f_{2\beta-2\alpha}$ and 
$f_{2\beta+2\alpha}$) 
in $M(N,H)$ as a function of $\widetilde{g}$ for 
$m_{\alpha}/m_{0}=0.5, m_{\beta}/m_{0}=1.0, \epsilon^{b}_{\alpha}=680$ and 
$\epsilon^{b}_{\beta}=0$. 
}
\end{figure}

It is clear from Fig. 4 that the $\widetilde{g}$-dependences 
of the FTAs are not described by 
the spin reduction factor for the single-band systems, 
eq. (4). 
For example, the 
FTAs of the second harmonics oscillations ($2\alpha$ and $2\beta$
oscillations) are not constant as a function of $\widetilde{g}$, 
which 
is not expected from eq. (4) for even $p$.

The FTAs of the fundamental oscillations ($\alpha$ and $\beta$ 
oscillations) become zero 
when $\widetilde{g}=\frac{1}{2}(m_0/m_i)(2n+1)$, 
which can be seen in Figs. 4 and 6. 
This 
is the same as the condition of the spin 
splitting zero given by eq. (4) for $p=1$. 

We can see from Figs 4 and 6 that 
the $\beta$+$\alpha$ and 
$\beta$-$\alpha$ oscillations are suppressed when 
the FTAs of $\alpha$ or $\beta$ oscillations becomes zero. 
This is in contrast with 
the magnetic breakdown 
model (Fig. 4 in ref.\cite{kishigispin}). For example, 
in the magnetic breakdown model 
the FTA of the $\beta$+$\alpha$ oscillation is large 
at $\widetilde{g}=0.4$, 
when the $\beta$ oscillation 
is suppressed. 
The large amplitude of the  $\beta$+$\alpha$ oscillation 
in the magnetic breakdown model might be understood as 
follows. 
The suppression of the $\beta$ oscillation is caused by 
the cancellation due to the 
$\pi$ phase difference between 
the oscillations for the up spin and 
the down spin in the semiclassical picture. 
This dose not result in the cancellation of 
the $\beta$+$\alpha$ oscillation 
in the magnetic breakdown systems 
because the $\beta$+$\alpha$ oscillation corresponds to the 
larger orbit caused by the tunneling and the phase difference 
between the up and down spins in the $\beta$+$\alpha$ orbit may 
not be $\pi$. 
In the two-band systems 
without the magnetic breakdown, 
the $\beta$+$\alpha$ oscillation is not caused by the larger 
orbit but the chemical potential oscillation. 
As a result, the $\beta$+$\alpha$ oscillation is 
suppressed in the two-band systems when $\alpha$ or $\beta$ oscillation 
is suppressed. 

The FTA of the $2\beta+2\alpha$ oscillation is 
constant as a function of $\widetilde{g}$, 
which can be seen in Figs. 5 and 7.

By tilting the magnetic field the 
suppression of the peak of the $\beta\pm\alpha$ oscillations 
and the constant peak of the $2\beta+2\alpha$ oscillation 
as a function of $\widetilde{g}$ 
may be observed in Sr$_2$RuO$_4$. 

The Yamaji effect\cite{yamaji} should be also considered 
when the magnetic field is tilted 
if the Fermi surface 
has weak three-dimensionality. 
Nakano\cite{nakanoya} shows that the $\beta\pm\alpha$ oscillations 
become large due to the Yamaji effect 
when $\alpha$ and $\beta$ oscillations
are enhanced in the spinless model. 
In quasi two-dimensional materials, 
the interplay between the spin effect studied in this 
paper and the Yamaji effect should be considered.

Ohmichi et al.\cite{ohmichi} have measured the FTAs of 
each oscillations in the magnetoresistance in Sr$_2$RuO$_4$, where 
there is no 
suppression of the peak of the $\beta\pm\alpha$ 
oscillations due to the spin 
although the enhancement of each oscillations due to Yamaji 
effect is seen. 
Our theory for the magnetization cannot be compared with 
their magnetoresistance experiment\cite{ohmichi}, 
because the Stark quantum 
interference oscillation\cite{fukuyama,stark} 
should be considered in 
the magnetoresistance. 
Yoshida et al.\cite{yoshida} have measured 
the angle-dependence of the FTA of 
the $\alpha$ oscillation. 
We expect that the effect of the spin and 
quasi-two-dimensionality can be observed in 
the angle-dependence of the FTAs of 
$\beta\pm\alpha$ and higher 
harmonic oscillations 
of the dHvA oscillation 
in Sr$_2$RuO$_4$. 


\begin{figure}
\leavevmode
\epsfxsize=8.5cm
\epsfbox{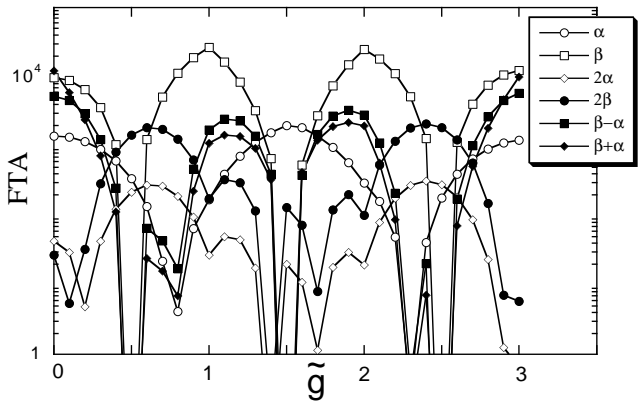}
\caption{The FTAs of 
some peaks ($f_\alpha$, $f_{2\alpha}$, $f_\beta$, $f_{2\beta}$, 
$f_{\beta-\alpha}$ and $f_{\beta+\alpha}$) 
in $M(N,H)$ as a function of $\widetilde{g}$ for  
$m_{\alpha}/m_{0}=0.65, m_{\beta}/m_{0}=1.0, \epsilon^{b}_{\alpha}=754$ and 
$\epsilon^{b}_{\beta}=0$. 
}
\end{figure}
\begin{figure}
\leavevmode
\epsfxsize=8.5cm
\epsfbox{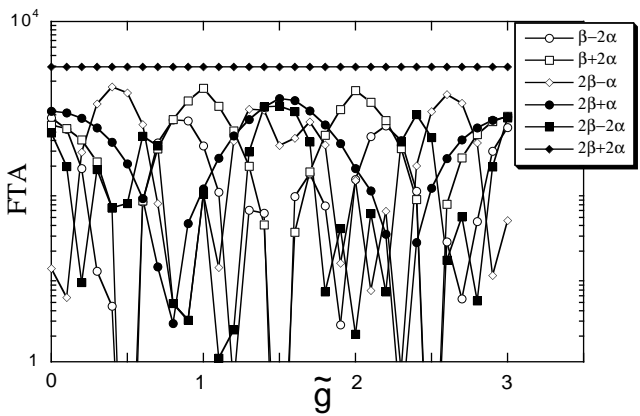}
\caption{The FTAs of 
some peaks ($f_{\beta-2\alpha}$, $f_{\beta+2\alpha}$, 
$f_{2\beta-\alpha}$, 
$f_{2\beta+\alpha}$, 
$f_{2\beta-2\alpha}$ and 
$f_{2\beta+2\alpha}$) 
in $M(N,H)$ as a function of $\widetilde{g}$ for 
$m_{\alpha}/m_{0}=0.65, m_{\beta}/m_{0}=1.0, \epsilon^{b}_{\alpha}=754$ and 
$\epsilon^{b}_{\beta}=0$. 
}
\end{figure}

In conclusion, we study the dHvA oscillation 
including the Zeeman effect for the two-dimensional 
two-band model, where the magnetic breakdown is 
neglected. 
We find the anomalous $\widetilde{g}$-dependences of the FTAs of 
$\beta$+$\alpha$, $\beta$-$\alpha$ and 2$\beta$+2$\alpha$ oscillations
in $M(N,H)$, where $\widetilde{g}=(gm_{i})/(2m_{0}\cos\theta)$. 
The $\beta\pm\alpha$ oscillations are suppressed when 
the $\alpha$ or $\beta$ oscillation 
disappears, and the FTA of the 2$\beta$+2$\alpha$ oscillation is 
constant as a function of 
$\widetilde{g}$. 
The effect of the spin on $M(N,H)$ cannot be described by 
the spin reduction factor for the single-band (eq. (4)), 
whereas that in $M(\mu,H)$ is given by eq.(2). 
Even in the two-band model, however, 
the spin splitting zero condition for fundamental 
frequencies ($\alpha$ and $\beta$ oscillations) 
in $M(N,H)$ is given by eq. (4). 
We expect that 
the $\widetilde{g}$-dependences of the amplitudes of these oscillations 
may be observed in the experiment of  
tilting magnetic field 
in two-dimensional multi-band system such as Sr$_2$RuO$_4$. 



We would like to thank 
M. Nakano for valuable discussions.
One of the authors (K. K.) was partially supported by Grant-in-Aid for
JSPS Fellows from the Ministry of
Education, Science, Sports and Culture.
K. K.
was financially supported by the Research Fellowships
of the Japan Society
for the Promotion of Science for Young Scientists.


\begin{thebibliography}{99}
\bibitem{FS}
Quasi-two-dimensional organic conductors 
($\kappa$-(BEDT-TTF)$_2$Cu(NCS)$_2$ and 
$\alpha$-(BEDT-TTF)$_2$KHg(SCN)$_4$) 
have the two energy bands as shown in Fig. 1(a). 
In quasi-two-dimensional material, Sr$_2$RuO$_4$, 
three closed orbits($\alpha$, $\beta$ and $\gamma$) exist. 
The contribution of the magnetic breakdown for 
the Fermi surface of Sr$_2$RuO$_4$ is much smaller than 
that of the quasi-two-dimensional organic 
conductors.
\end{thebibliography}
\end{document}